\documentclass[amsmath,amssymb,pra,twocolumn,showpacs]{revtex4}
\usepackage[cp1251]{inputenc}
\usepackage[english]{babel}
\usepackage{color}
\usepackage{graphicx}
\usepackage{subfigure}
\usepackage{epstopdf}
\usepackage{dcolumn}
\usepackage{tabularx}
\usepackage{array}
\usepackage{braket}
\newcommand{\abs}[1]{\left| #1 \right|}
\usepackage{amsmath}
\usepackage{natbib}
\numberwithin{equation}{section}

\begin{document}
\bibliographystyle{unsrt}

\title{Spectroscopy of the atomic system driven by high intensity laser field}


\author{A.V. Bogatskaya$^{1, 3}$}
\email{annabogatskaya@gmail.com}
\author{E.A. Volkova$^{1}$}
\author{A.M. Popov$^{1, 2, 3}$}
\email{alexander.m.popov@gmail.com}
\address{$^1$D.V. Skobeltsyn Insititute of Nuclear Physics, Moscow State University, Leninskie Gory, 119991, Moscow Russia \\
$^2$Department of Physics, Moscow State University, Leninskie Gory, 119991, Moscow Russia \\
$^3$P.N. Lebedev Physical Institute, Leninskii prospekt, 53, 119991, Moscow, Russia}

\date{ }

\begin{abstract}
Spontaneous emission of the quantum system driven by a high intensity classical laser field is analyzed. The study is based on the accurate consideration of quantum system interaction with vacuum quantized field modes in the first order of perturbation theory, while the intense laser field is considered classically beyond the perturbation theory. It is demonstrated that the spectrum of the spontaneous emission can be used for analyzing of the strong-field dynamics and structure of the energy spectrum of the atomic system. The obtained data are compared with that obtained in the frames of semiclassical approximation typically used for analyzing of the strong-field dynamics. It is found that the applicability of the semiclassical approach is strictly limited.

\end{abstract}

\maketitle
\bibliographystyle{unsrt}

\section{Introduction}
Analysis of the strong field atomic dynamics in one of the most important problems of modern atomic physics and extreme nonlinear optics \cite{agostini2004physics, krausz2009attosecond}. It is known, that atomic spectrum can be dramatically reconstructed  by an intense external laser field. The new so-called "dressed atom" appears to exist  \cite{delone2012multiphoton,fedorov1997atomic}. It can be named as an example the Kramers - Henneberger atom arising in the adiabatic stabilization regime in a strong laser field \cite{gavrila1984free,pont1988dichotomy,pont1990stabilization}. Another kind of dressing is realized in the interference stabilization (IS) regime \cite{fedorov1988field,fedorov1990coherence,fedorov1998strong}. The detailed analysis of the stabilization phenomenon can be found in \cite{fedorov1997atomic} and reviews \cite{gavrila2002atomic,popov2003strong}. We would like to note that significant dressing and reconstruction of the atomic spectrum can also appear in rather weak external fields if one studies the resonant interaction of the laser field with an atom. The example of such a dressing is splitting of atomic energy levels in the process of Rabi oscillations and formation of the Mollow triplet \cite{mollow1969power}. Spontaneous emission of such dressed by external fields atoms is one of the best ways to study the structure of energy spectra arising during the dressing. 

On the other hand, to study the initial stage of any nonlinear process the spontaneous emission should be also taken into account. Typically the nonlinear response is analyzed by the semiclassical approach \cite{krause1992high,l1993high,lewenstein1994theory,becker1994modeling,platonenko1998generation} when electromagnetic field is considered still classically. To introduce spontaneous emission to the semiclassical theory the correspondence principle \cite{fedorov1997atomic} is applied. Such an approach was questioned in \cite{bogatskaya2016polarization,bogatskaya2016spontaneous}. It was demonstrated that the application of the semiclassical approach to study emission of the quantum system driven by high intensity laser field is generally in contradiction with quantum electrodynamical calculations. 

In this paper new approach for analysing spontaneous radiative processes in a quantum system driven by a strong classical laser field is considered. This approach is based on the first order perturbation theory applied to the interaction of the atomic system dressed by the external laser field with a lot of quantized field modes in the assumption that initially all the modes are in a vacuum state. The proposed approach is applied to study the spectrum and emission of a model single atom driven by the strong laser field, in particular in the regime of the strong-field stabilization. Comparison of the obtained data with that obtained in the semiclassical approximation is performed.

\section{Atomic dynamics in a strong laser field in the presence of interaction with quantized electromagnetic fields}

We analyse the atomic system using the following Hamiltonian
	\begin{equation}
	H=H_{0}(\vec{r},t)+H_f(\lbrace a \rbrace)+V(\vec{r},\lbrace a \rbrace),
 	\label{Hamilt}
	\end{equation}
where $H_0=H_{at}(\vec{r})+W(\vec{r},t)$, $H_{at}(\vec{r})$ is the atomic Hamiltonian,
 	\begin{equation}
	W=-\frac{e}{mc}\vec{A}(t) \vec{p}+\frac{e^{2} A^{2}(t)}{2mc^{2}}
 	\label{W}
	\end{equation}
represents the interaction of atom with classical laser field in the velocity gauge in the dipole approximation, $\vec{A}(t)$ is the vector-potential of the classical field, $\vec{p}=-i\hbar\nabla$ is the momentum operator, $H_f$ is the Hamiltonian of the set of field modes excluding laser field mode, $V$ stands for the interaction of an atomic electron with the quantized electromagnetic field, $\vec{r}$ is the electron radius-vector and $\lbrace a\rbrace$ is the set of quantized field mode coordinates.
We are going to deal with the quantized field using perturbation theory. Then, in the case when there is no interaction one can write the well-known equation
     \begin{equation}
	i\hbar\frac{\partial\psi(\vec{r},t)}{\partial t}=H_0(t)\psi(\vec{r},t),
	 \label{Schroed}
	\end{equation}
which describes the atom dynamics in a classical laser field. Initial condition can be written as 
	 \begin{equation}
	\psi(\vec{r},t=0)=\phi(\vec{r}),
	 \label{Initial}
	\end{equation}
where $\phi(\vec{r})$ is a given stationary or unstationary state of the atomic discrete spectrum or continuum. We will also suppose  that at the initial instant of time all field modes are in the vacuum state $\ket{\lbrace0 \rbrace}$. Provided that we know the solution of equation (\ref{Schroed}), the solution of the general equation with the Hamiltonian (\ref{Hamilt})
	 \begin{equation}
	i\hbar\frac{\partial\psi(\vec{r},\lbrace a \rbrace,t)}{\partial t}=(H_0(t)+ H_f + V)\psi(\vec{r},\lbrace a \rbrace,t),
	\label{Schroed2}
	\end{equation}
and initial condition $\psi(\vec{r},\lbrace a \rbrace,t=0)=\phi(\vec{r})\times \ket{\lbrace 0 \rbrace}$ can be found by means of perturbation theory.

Wave function of zero-order approximation excluding interaction with the quantum field modes reads
	\begin{equation}
	\Psi^{(0)}(\vec{r},\lbrace a \rbrace,t)=\psi(\vec{r},t)\times \ket{\lbrace 0 \rbrace}.
	\label{zeroorder}
	\end{equation}
We are going to find the solution of (\ref{Schroed}) in the form:
	\begin{equation}
	\Psi(\vec{r},\lbrace a \rbrace,t)=\Psi^{(0)}(\vec{r},\lbrace a \rbrace,t)+ \delta \Psi(\vec{r},\lbrace a \rbrace,t)
    \label{solutionform}
	\end{equation} 
assuming that $\delta\Psi \ll \Psi^{(0)}$.

Substituting (\ref{solutionform}) in (\ref{Schroed2}) in the first order of smallness one obtains:
	\begin{gather}
 	\nonumber
 	i\hbar\frac{\partial\delta \Psi(\vec{r},\lbrace a \rbrace,t)}{\partial t}=(H_0(t)+ H_f)\delta\Psi(\vec{r},\lbrace a \rbrace,t)+\\ 
    V \Psi^{(0)}(\vec{r},\lbrace a \rbrace,t),
 	\label{Schroed3}
	\end{gather}
 with the initial condition $\delta \Psi(\vec{r},\lbrace a \rbrace,t=0)=0$.

In fact (\ref{Schroed3}) can be formulated as Schroedinger equation for the $\delta \Psi$ with the source in the right hand. For further analysis of eq. (\ref{Schroed3}) let us remind that initially we have vacuum in all field modes. Therefore in the first order of perturbation theory $\delta \Psi$ contains only one-photon excitation in a some field mode: 
	\begin{equation}
	\delta \Psi(\vec{r},\lbrace a \rbrace,t)=\sum\limits_{k,\lambda}\delta \psi_{k\lambda}(\vec{r},t)\times \lbrace 0,0,\ldots 1_{k,\lambda},0,\ldots0,0 \rbrace.
    \label{deltapsi}
	\end{equation} 
Here $\delta \psi_{k\lambda}(\vec{r},t)$ is the electron wave function provided that one photon with wave vector $\vec{k}$ and polarization $\lambda$ has appeared.
As the interaction of the atom with quantized field can be written in a form

	\begin{equation}
	V(\vec{r},\lbrace a \rbrace)=\sum\limits_{k,\lambda}V_{k \lambda}=-\frac{e}{mc}\sum\limits_{k,\lambda}(\vec{e}_{k \lambda} \vec{p})a_{k \lambda}
    \label{interaction_with_Qmodes}
	\end{equation} 
( $a_{k \lambda}$ is the vector-potential operator of mode $\lbrace{k, \lambda}\rbrace$  and $\vec{e}_{k \lambda}$ is the polarization vector) for a given mode with one-photon excitation one can write:
	\begin{gather}
 	\nonumber
 	i\hbar\frac{\partial\delta \psi_{k\lambda}(\vec{r},t)}{\partial t}\ket{1_{k\lambda}} + \delta \psi_{k \lambda}(\vec{r},t)\times (3/2)\hbar \omega_{k \lambda}\ket{1_{k\lambda}}=\\
 	\left( H_0(t)+ h^{(f)}_{k \lambda}\right)\delta\psi_{k\lambda}(\vec{r},t)\ket{1_{k\lambda}}+V_{k \lambda}\psi(\vec{r},t)\ket{0}.
 	\label{onephot_exitation_eq}
	\end{gather}
Here $h^{(f)}_{k \lambda}$ is the Hamiltonian of the field mode $\lbrace{k, \lambda}\rbrace$. Taking into account that $h^{(f)}_{k\lambda} \ket{1_{k\lambda}}=3/2\cdot \hbar\omega_{k \lambda}\ket{1_{k\lambda}}$, and $a_{k \lambda}\ket{0}=\frac{a_{norm \,k \lambda}}{\sqrt{2}}\ket{1_{k\lambda}}\exp(i\omega_{k\lambda}t)$, ($a_{norm \,k \lambda}=\sqrt{4\pi\hbar c^{2}/ \omega_{k \lambda}L^{3}}$) is the auxiliary normalizing constant, $L$ is normalization volume), the final form of the eq. (\ref{onephot_exitation_eq}) can be written as
	\begin{gather}
 	\nonumber
 	i\hbar\frac{\partial\delta \psi_{k\lambda}(\vec{r},t)}{\partial t}=H_0\delta \psi_{k \lambda}(\vec{r},t)\\
 	-\frac{e(\vec{e}_{k \lambda}\vec{p})}{mc} \times \frac{a_{norm \,k \lambda}}{\sqrt{2}}\times \psi(\vec{r},t) \times \exp(i\omega_{k \lambda}t)
 	\label{main_Schroed}
	\end{gather}
with the initial condition $\delta\psi_{k \lambda}(\vec{r},t=0)=0$.

It is obvious, that the expression 
	\begin{equation}
	W_{k \lambda}(t)=\int|\delta\psi_{k \lambda}(\vec{r},t)|^{2}d^{3}r
    \label{probability}
	\end{equation}
represents the probability to find a photon in the mode $\ket{k,\lambda}$ as a function of time. Then the total probability to emit the photon of any frequency and polarization during the transition $f\rightarrow i$ is 
	\begin{equation}
	W_{fi}(t)=\sum\limits_{k,\lambda}W_{k \lambda}(t).
    \label{totalprobability}
	\end{equation}
As the spectrum of field modes is dense, we can replace the sum in eq. (\ref{totalprobability}) by the integral:
	 \begin{equation}
	2L^{3}\int\frac{d^{3}k}{(2\pi)^{3}}=\dfrac{2L^{3}}{8\pi^{3}c^{3}} \int \omega^{2}_{k \lambda}d \omega_{k \lambda} \int d \Omega.
    \label{totalprobabilityintegr}
	\end{equation}
After the integration over angular distribution of photons and summation over possible polarizations the probability of the spontaneous decay in the spectral interval $(\omega, \omega+d \omega)$ can expressed in the form
	\begin{equation}
	W_{\omega}d \omega= \frac{L^{3}}{3\pi^{2}c^{3}}\omega^{2}d \omega \times W_{k=\omega/c,\lambda},
    \label{int_probab_w}
	\end{equation}
where $W_{k \lambda}$  is given by (\ref{probability}). One should note, that the expression (\ref{int_probab_w}) do not depend on the normalization volume, as $W_{k \lambda} \sim 1/L^{3}$. The energy emitted in the spectral interval $(\omega, \omega+d \omega)$ is found as:
	\begin{equation}
	dQ_\omega=\hbar\omega \times W_{\omega}d \omega.
	\end{equation}
	
	To provide more insight into the physics of spontaneous emission in the presence of strong laser field, wave functions $\delta\psi_{k\lambda}(\vec{r},t)$ should be represented as the superposition of the stationary states of the atomic Hamiltonian: 
	\begin{equation}
	\delta\psi_{k\lambda}(\vec{r},t)=\sum\limits_{n}C_{n}^{(k \lambda)}(t)\varphi_{n}(\vec{r})\exp\left( -\frac{i}{\hbar}E_{n}t\right),
	\label{decomposing}
	\end{equation}
The squared coefficients of the decomposition (\ref{decomposing}) imply the probability to find the atom in different states under the assumption that the emitted photon is in the definite mode $\lbrace k,\lambda \rbrace$.  We will further use these values to interpret the results of numerical simulation.

We would like to mention that the discussed problem can be formulated also in the length gauge   \cite{bogatskaya2016spontaneous}. In this case 
	\begin{equation}
	W=-\vec{d}\vec{\varepsilon}(t),
	\end{equation}
where $\vec{d}$ is the dipole operator and $\vec{\varepsilon}(t)=-\frac{1}{c}(d \vec{A}/dt)$ is electric field strength. However, it is known \cite{fedorov1997atomic} that if the dipole approximation is valid, both gauges provide equivalent results if we deal with the exact solution of the nonstationary problem (\ref{Schroed}).

\section{Spontaneous Raman and Rayleigh scattering}

First, we will start from the case when the solution of nonstationary Shroedinger equation (\ref{Schroed}) for the atomic system driven by the classical laser field can be found approximately in the first order of perturbation theory. Then
	\begin{gather}
	\nonumber{}
	\psi(\vec{r},t)=\varphi_i(\vec{r})\exp \left(-\frac{i}{\hbar}E_{i}t \right)+\\
	\sum\limits_{n\neq i}C_{n}^{(1)}(t)\varphi_n(\vec{r})\exp \left(-\frac{i}{\hbar}E_{n}t \right),
	\label{perturb_theory_decomposing}
	\end{gather}
where 
 	\begin{equation}
	C_{n}^{(1)}(t)=\frac{\vec{d_{ni}}\vec{\varepsilon_0}}{2 \hbar}\left(\frac{\exp(i(\omega_{ni}-\omega)t)}{(\omega_{ni}-\omega)}+\frac{\exp(i(\omega_{ni}+\omega)t)}{(\omega_{ni}+\omega)} \right).
	\label{C_n1}
	\end{equation}
Here $\varphi_{n}(\vec{r})$ and $E_{n}$ are the eigenstates and eigenvalues of the atomic Hamiltonian $H_{at}(\vec{r})$, and $\omega_{ni}$ are the transition frequencies from state $\ket{i}$ to state $\ket{n}$. We use here the length gauge as far as for rather weak monochromatic (quasi-monochromatic) field   $\vec{\varepsilon}=\vec{\varepsilon}_{0}\cos(\omega t)$ the relation $\vec{p}_{ni}\vec{A}_{0}=\vec{d}_{ni}\vec{\varepsilon}_{0}$ is obviously valid and both gauges provide the same result. Substituting (\ref{perturb_theory_decomposing}) and (\ref{C_n1}) into (\ref{main_Schroed}) and assuming that $\ket{i}$ is the initial atomic state one derives the equation for probability amplitude to find the atom in $\ket{f}$ and the photon in the mode $\lbrace k, \lambda \rbrace$: 
\begin{gather}
	\nonumber{}
	i\hbar\psi\dot{C}_{f}^{(k\lambda)}= -\sum_{n}{C}_{n}^{(k\lambda)}(\vec{d}_{fn}\vec{E}(t))\exp(i\omega_{fn}t)-\\
	\frac{\varepsilon_{norm}}{\sqrt{2}}\sum_{n \neq i}C_{n}^{(1)}(\vec{e}_{k\lambda}\vec{d}_{fn})\exp(i(\omega_{fn}+\omega_{k\lambda})t),
	\label{perturb_theory_equation}
	\end{gather}
Here $\varepsilon_{norm\,k\lambda}=\omega_{k \lambda}a_{norm\,k\lambda}/c=\sqrt{4\pi\hbar\omega_{k\lambda}/L^{3}}$ is the normalizing constant for the electric field in the mode $\lbrace k, \lambda \rbrace$.
Second term in the right part of (\ref{perturb_theory_equation}) stands for the emission of photon $\lbrace k, \lambda \rbrace$ while the first one describes the evolution of atomic wave function in the classical field after the emission of photon and here we will neglect such evolution. From (\ref{perturb_theory_equation}) one derives:
	\begin{gather}
	\nonumber{}
	|C_{f}^{k \lambda}(t \to \infty)|^2=\varepsilon_{norm\,k\lambda}^2 \left| \sum_{n \neq i}\frac{(\vec{d}_{fn}\vec{e}_{k \lambda})(\vec{d}_{ni}\vec{\varepsilon}_{0})}{2 \sqrt{2}\hbar}\frac{1}{\omega_{ni}-\omega}\right|^2 \\
	\times 2\pi\delta(E_{f}-E_{i}-\hbar\omega+\hbar\omega_{k \lambda})\times t,
	\label{raman_scatt}
	\end{gather}
This is the probability of the Stoks component of the spontaneous Raman scattering corresponding to the final state $\ket{f}$ and emission of a photon
	\begin{equation}
	\hbar\omega_{k\lambda}=\hbar\omega-(E_{f}-E_{i}).
	\label{}
	\end{equation}
If the final state coincides with the initial one $\ket{f}=\ket{i}$ we derive the expression for the Rayleigh scattering
	\begin{gather}
	\nonumber{}
	\left|C_{i}^{k \lambda}(t \to \infty)\right|^2=\varepsilon_{norm\,k\lambda}^2 \left| \sum_{n \neq i}\frac{(\vec{d}_{in}\vec{e}_{k \lambda})(\vec{d}_{ni}\vec{\varepsilon}_{0})}{2 \sqrt{2}\hbar}\frac{1}{\omega_{ni}-\omega}\right|^2 \\
	\times 2\pi\delta(\hbar\omega_{k \lambda}-\hbar\omega)\times t,
	\label{rayleigh_scatt}
	\end{gather}
when the frequency of spontaneous photon is the same as for laser radiation. Not far from the resonance when the laser frequency is $\omega\approx\omega_{ni}$ with definite value of $n$ , the transition takes place through the only intermediate state and summation over all intermediate states in expressions (\ref{raman_scatt}) and (\ref{rayleigh_scatt}) should be omitted: 
 	\begin{gather}
	\nonumber{}
	\left|C_{i}^{k \lambda}(t \to \infty)\right|^2=\varepsilon_{norm\,k\lambda}^2 \left|\frac{(\vec{d}_{in}\vec{e}_{k \lambda})(\vec{d}_{ni}\vec{\varepsilon}_{0})}{2 \sqrt{2}\hbar}\frac{1}{\omega_{ni}-\omega}\right|^2 \\
	\times 2\pi\delta(E_{f(i)}-E_{i}-\hbar\omega+\hbar\omega_{k \lambda})\times t,
	\label{resonant_scatt}
	\end{gather}
In higher orders of perturbation theory for interaction with the classical laser field it is possible to obtain similar expressions for hyper-Rayleigh $$\hbar\omega_{k\lambda}=n\hbar\omega,\,\,n=2,3,4,\dots $$
and hyper-Raman scattering of laser radiation by atomic system $$\hbar\omega_{k\lambda}=n\hbar\omega-(E_{f}-E_{i}),\,\,n=2,3,4,\dots $$
For example, for Stocks hyper-Raman scattering with $n=2$ one reads
\begin{gather}
\nonumber{}
\abs{C^{k\lambda}_i(t \leftarrow \infty)}^2 = -\frac{\varepsilon^2_{norm}}{2\hbar} \Bigg |\sum_{m \neq i} (\vec e_{k \lambda} \vec d_{fm}) \times\\
\nonumber \sum_n \frac{(\vec d_{mn}\vec \varepsilon_0)(\vec d_{ni}\vec \varepsilon_0)}{4\hbar^2} \times \frac{1}{(\omega_{mi}-\omega)(\omega_{ni}-\omega)} \Bigg |^2 \times \\
2\pi \delta (E_f-E_i-2\hbar \omega +\hbar \omega_{k \lambda})
\end{gather}

\section{Numerical simulation of spontaneous emission in the presence of strong laser field}	
 In this section we briefly describe the numerical procedure that was used to study the spontaneous emission of the quantum system driven by high-intensity laser field.

We study one-dimensional atomic system with the Coulomb-screened potential \cite{javanainen1988numerical}:
	\begin{equation}
	V(x)=-\frac{e^2}{\sqrt{\alpha^2+x^2}}.
	\label{Coulomb-sc potential}
	\end{equation}
Screening parameter is $\alpha=1.6165\,a_0$, $a_0$ is Bohr radius. For such value of $\alpha$ the ionization potential is $12.13$ eV which corresponds to ionization potential of xenon atom. The set of energy levels in such a xenon-like atom can be found in Table \ref{Table 1}.

\begin{table}[h]
\caption{Xenon energy levels obtained in the numerical simulations for the potential (\ref{Coulomb-sc potential})} 
\centering
\begin{tabularx}{\linewidth}{>{\centering}p{6cm}<{\centering}p{2.5cm} }
\hline
Principal quantum number, n &    Energy, eV \\ \hline
1 & -12.134 \\
2 & -5.910 \\
3 & -3.457 \\
4 & -2.220 \\
5 & -1.550 \\
6 & -1.135 \\
7 & -0.870 \\
8 & -0.685 \\
9 & -0.555 \\
10 & -0.457 \\
11 & -0.385 \\
12 & -0.327 \\
13 & -0.282 \\
\hline
\label{Table 1}
\end{tabularx}
\end{table}
Further we discuss the set of calculations for an atom being initially prepared in the ground state $n=1$ and exposed to the radiation of the second harmonic of Ti-Sa laser ($\hbar\omega=3.1$ eV) with the sine-squared pulse form:
\begin{equation}
	\varepsilon(t)=\varepsilon_0\sin^2{\frac{\pi t}{2 t_p}}\cos\omega t,
	\end{equation}
The half-pulse duration $t_p$ is 12 optical cycles. According to the above-mentioned model we solve self-consistently equation (\ref{Schroed}) for the evolution of atomic wave function with the set of equations (\ref{main_Schroed}) for one-photon excitations in different quantized field modes. The system of equations (\ref{Schroed}), (\ref{main_Schroed}) was solved numerically by the method of finite elements \cite{fletcher1984computational}. Main features of applied algorithm were discussed in \cite{popov2011numerical}. The number of modes considered in the calculations was 500-1000. Frequency interval between two neighboring modes was chosen about $0.02-0.04$ eV. The modelling was performed for the time interval twice exceeding the duration of pulse. That allows one to distinguish spontaneous transitions that are possible without laser field from the stimulated transitions like Raman and Rayleigh type when spontaneous photons appear only during the laser pulse action.

\begin{figure}[h]
\centering\includegraphics[width=8 cm]{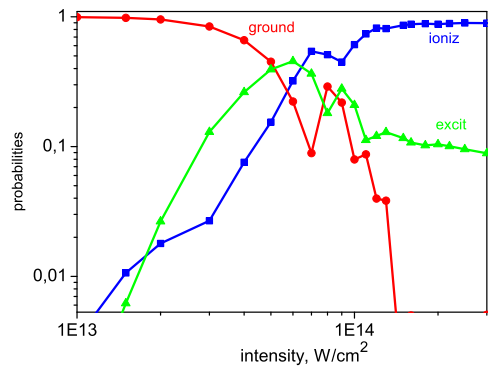}
\caption{{\protect\footnotesize {Ionization yield, probabilities of excitation of Rydber states and population of initial (ground) state in dependence on the intensity of laser radiation.}}}\label{Fig1}
\end{figure}

We will start from the discussion of the ionization dynamics of the ground state of the model atom ($n=1$). The obtained data for probabilities of ionization, excitation of Rydberg\textsuperscript{\footnotemark[1]}
\footnotetext[1]{Here we suppose that Rydberg states are the all states that bound with the continuum by single-photon transitions} states  as well as for population of the initial (ground) state are presented at fig.\ref{Fig1}. First we would like to note that ionization yield is found to be the nonmonotonous function of laser intensity as far as significant part of population is trapped in excited (Rydberg) states and can be interpreted as the interference stabilization phenomenon \cite{fedorov1988field,fedorov1990coherence,fedorov1998strong,fedorov2012interference}. Such behavior of the atomic system results from the multiphoton coupling of the ground states and set of Rydberg states near the continuum boundary and further population trapping (stabilization) in Rydberg states. Such situation was studied recently in detail  in a number of papers \cite{fedorov2012interference,popov2010low,volkova2011ionization}. Another point of view on the same problem was suggested in \cite{nubbemeyer2008strong,eichmann2013observing,zimmermann2016atomic}, where the model of frustrated tunneling was developed. In \cite{popov2014population} it was supposed that both models are similar, but one of them explains the process of atomic ionization in field-free atomic basis while another one deals with the dressed basis reconstructed by the laser field. 
\begin{figure}
\centering\includegraphics[width=8 cm]{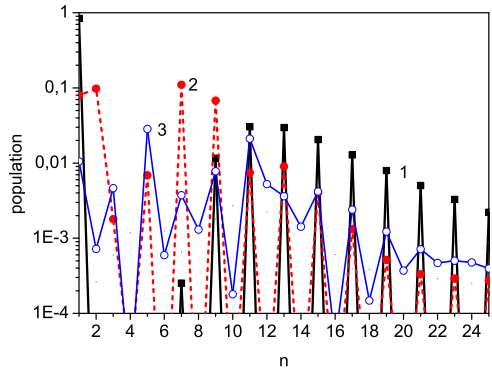}
\caption{{\protect\footnotesize {The population distribution over principal quantum number by the end of the laser pulse for intensities (W/cm$^2$) 1 - $3\times 10^{13}$, 2 - $10^{14}$, 3 - $3\times 10^{14}$ W/cm$^2$.}}}\label{Fig2}
\end{figure}

As we have already mentioned, efficient population of Rydberg states in high-intensity limit is the main feature of stabilization phenomenon. The distribution of excited states over principal quantum number for different peak laser pulse intensities is presented at fig.\ref{Fig2}. It can be seen that significant part of the atomic population is distributed over Rydberg states (with principal quantum numbers $n\geq 5$) with even values of quantum number corresponding to the four-photon transition from the ground state. Such population can be considered as a manifestation of the Raman $\Lambda$-type transitions via the continuum that testifies the phenomenon of interference stabilization [8].
\begin{figure}
\centering\includegraphics[width=8 cm]{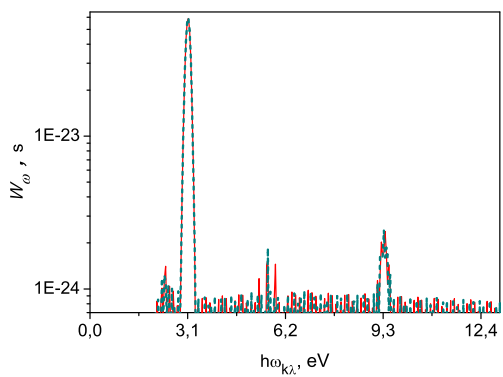}
\caption{{\protect\footnotesize {Spectral probability density of the atomic system emission irradiated by the $2^{nd}$ harmonic of Ti-Sa laser pulse. The intensity of radiation is $10^{13}$ W/cm$^2$ (c). Solid curve is the emission spectrum by the end of the laser pulse, dash curve corresponds to the instant of time equal two pulse duration.}}}\label{Fig3}
\end{figure}

Recent experiments \cite{zimmermann2017unified} on excitation and ionization dynamics of Ar and Ne atoms exposed to strong femtosecond laser pulses at $\lambda=400$ and $800$ nm have also demonstrated non-monotonous atom excitation yield dependence on laser intensity as a result of channel closing and effective population trapping in Rydberg states. The obtained data are found to be in qualitative agreement with theoretical model \cite{fedorov2012interference,popov2010low,volkova2011ionization}.

Calculations of the spectral probability density of photon emission $W_\omega$ from the atomic system driven by the femtosecond laser pulse for the relatively low intensity ($10^{13}$ W/cm$^2$) when the system dominantly stays in the initial ground state (see fig.\ref{Fig1}) are displayed at the figure \ref{Fig3}. Two peaks in the emission spectrum are well pronounced. In both cases the emission occurs only during the laser pulse action. The frequency of one of them coincides with the frequency of laser radiation $\hbar\omega_{k\lambda}=\hbar\omega$ and thus this peak represents the well known Rayleigh scattering. Another one corresponds to the third harmonic of laser radiation and can result from the hyper-Rayleigh scattering  $\ket{1}+3\hbar\omega \to \ket{1}+\hbar\omega_{k\lambda}$, $\hbar\omega_{k\lambda}=3\hbar\omega$. This process is responsible for the cubic nonlinearity of the media and third-order harmonic generation. Nevertheless, in our case the emission line at $\hbar\omega_{k\lambda}=3\hbar\omega=9.3$ eV can also have another origin. Really, the calculation of probabilities $\left| C_{n}^{k\lambda} \right|^2$ to find the atom in different bound states $\ket{n}$ under the assumption that a photon is in the mode $\lbrace k,\lambda\rbrace$ with $\hbar\omega_{k\lambda}=9.3$ eV demonstrates that atom is found to be in the state $\ket{n=1}$ with probability $\approx 0.64$, while the population of $\ket{n=2}$ is $\approx 0.36$. Such situation can be interpreted as a hyper-Raman scattering  $\ket{1}+5\hbar\omega \to \ket{2}+\hbar\omega_{k\lambda}$ via intermediate continuum state. 

We would also like to note that as the stationary states in discrete spectrum of the atomic system are characterized by the definite parity, the second harmonic and other even harmonics in the emission spectrum are forbidden.

\begin{figure}[!h]  
\vspace{-4ex} \centering \subfigure[]{
\includegraphics[width=\linewidth]{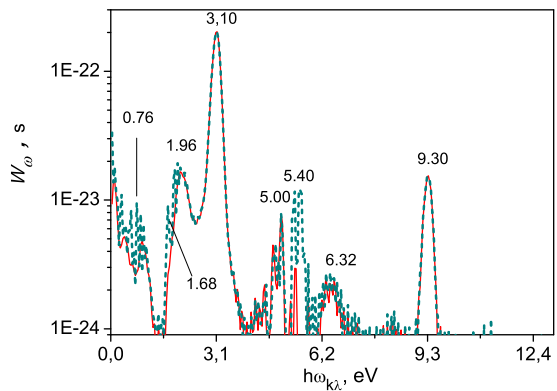} \label{Fig4a} }  
\hspace{4ex}
\subfigure[]{
\includegraphics[width=\linewidth]{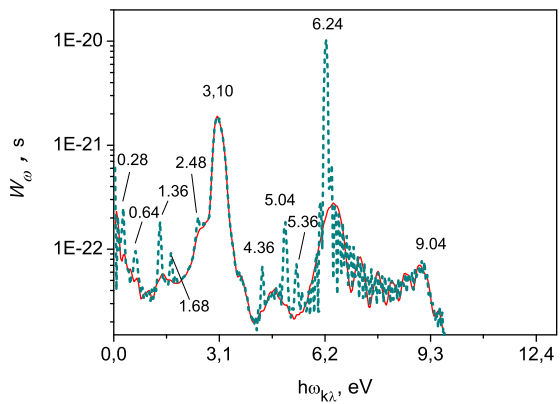} \label{Fig4b} }
\hspace{4ex}
\subfigure[]{\includegraphics[width=\linewidth]{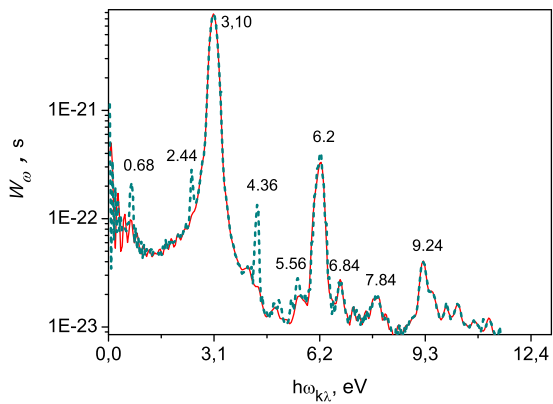} \label{Fig4c}}  
\caption{Similar to fig.\ref{Fig3}, but for laser intensities (in W/cm$^2$):  $3\times 10^{13}$ \subref{Fig4a}, $10^{14}$ \subref{Fig4b} and $3\times 10^{14}$ \subref{Fig4b}. Values near the peaks are their energy positions (in eV). Solid curve is the emission spectrum by the end of the laser pulse, dash curve corresponds to the instant of time equal two pulse duration.} \label{Fig4}
\end{figure}

For higher intensities values when the essential excitation and ionization of the atom takes place, the emission spectra are much more complicated (see fig.\ref{Fig4}). Besides above mentioned lines corresponding to the fundamental frequency and it's third harmonics, a lot of other peaks are found to be seen. Some of them are formed during the laser pulse action while others appear in the after-pulse regime.
 
Nevertheless, for high intensities even the peak with $\hbar\omega_{k\lambda}=\hbar\omega$ can have several physical origins. One of them which has  been already mentioned when a first stimulated one-laser-photon transition from the initial state to the intermediate state is followed by a spontaneous transition from this state back to the initial state. Another one is the bremsstrahlung transition at the radiation frequency appearing from the free electron oscillation nearby the area of atomic potential. To distinguish these mechanisms the distribution of probabilities $\left| C_{n}^{(k\lambda)} \right|^2$ under the assumption that photon is in the mode $\lbrace k,\lambda\rbrace$ with  $\hbar\omega=3.1$ eV was also calculated. If the main contribution to the peak at $\hbar\omega_{k\lambda}=3.1$ eV appears from the Rayleigh scattering the corresponding value of $\left| C_{n=1}^{(k\lambda)} \right|^2$ will be close to unity. Such situation is really realized for intensities $\leq3 \times 10^{13}$ W/cm$^2$ while for higher intensities the main contribution comes from the bremsstrahlung effect. The same conclusion is also valid for the peak with $\hbar\omega_{k\lambda}=9.1$ eV. In both cases this emission takes place only during the laser pulse action.

Very interesting situation is realized for the peak with frequency near $6.2$ eV that appears to exist in the strong field limit $I\geq 10^{14}$ W/cm$^2$. First, the second harmonic frequency of the laser pulse which should be forbidden by selection rules for the case of atom being initially in the ground state is observed here. Such an emission line can be associated with transitions between different continuum states (see fig.\ref{Fig4c}). On the other hand the discussed frequency is close to the frequency of the transition $\hbar\omega_{21}=E_2-E_1\approx6.22$ eV. Contribution of such spontaneous transition in the after-pulse regime is clearly seen: calculation of the coefficient $\left| C_{n=1}^{(k\lambda)} \right|^2$ for the $\hbar\omega_{k\lambda}=6.2$ eV gives the value $\sim1$ for the intensity $10^{14}$ W/cm$^2$.
    
     We can distinguish two groups among other lines: one of them appears in after-pulse regime while the second one forms during the laser pulse action. Really, lines with $\hbar\omega_{k\lambda}=4.36,5.04,5.36,5.56$ eV correspond to the spontaneous transitions to the state $\ket{2}$ from upper Rydberg states $n=5, 7, 9, 11$ correspondingly. Lines with $\hbar\omega_{k\lambda}=0.68,1.36,1.66$ eV correspond to the spontaneous transitions to the state $\ket{4}$ from states $n=5, 7, 9$. Line with $\hbar\omega_{k\lambda}=0.28$ eV can be interpreted as the head line of series $\ket{n}\to \ket{6}$ with $n=7$. 
 \begin{figure}
\centering\includegraphics[width=8 cm]{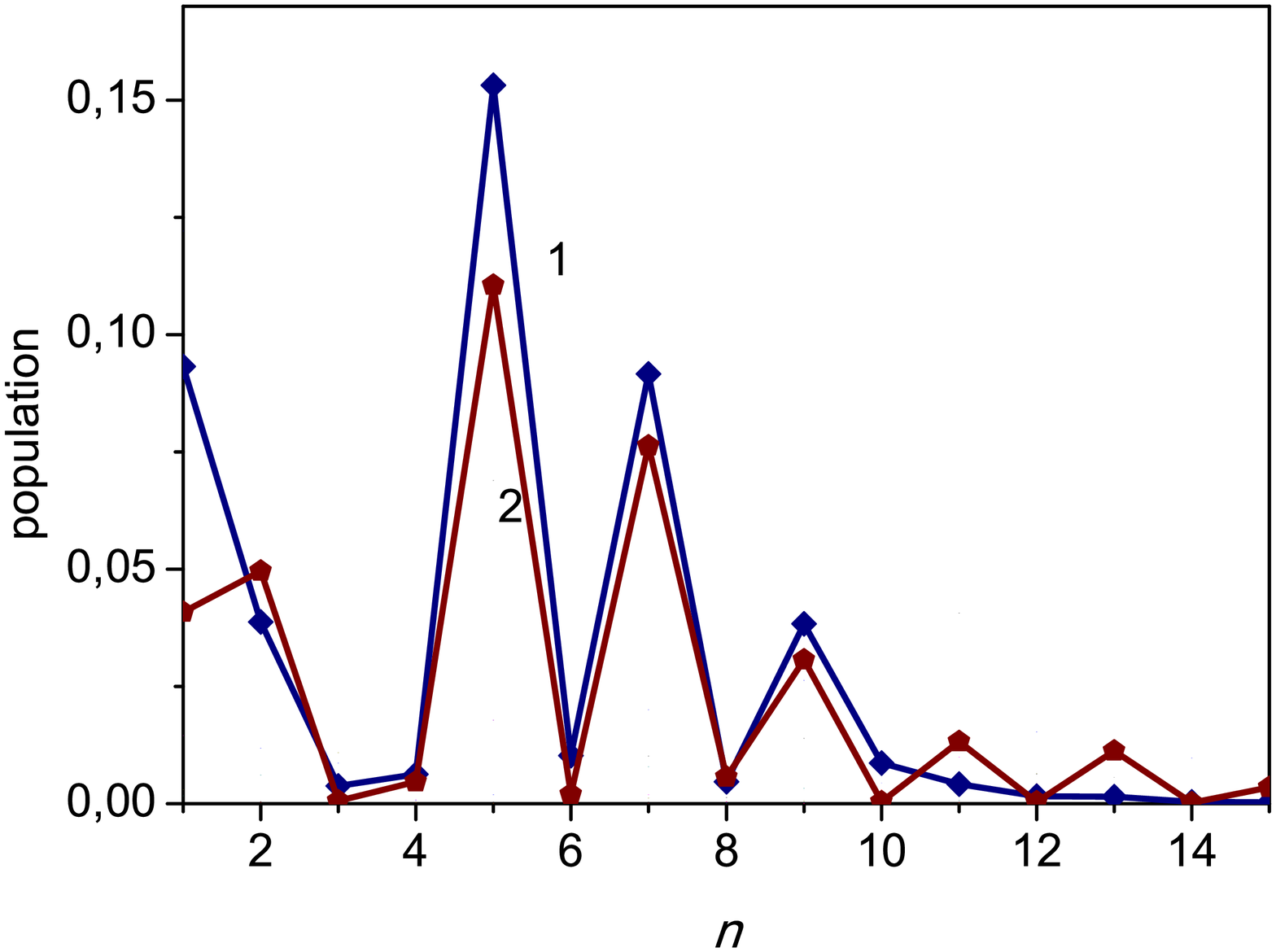}
\caption{{\protect\footnotesize {Population of different atomic states for the spontaneous emission of quanta $6.84$ eV (1) and $7.84$ eV (2) in the presence of a strong laser field with peak intensity $3\times10^{14}$ W/cm$^2$.}}}\label{Fig5}
\end{figure}

As above discussed structure of the spontaneous emission spectrum formed in the after-pulse regime, obtained data can be considered as the manifestation of population trapping in Rydberg states as a result of stabilization phenomenon.
    
    Now we will briefly discuss several peaks $\hbar\omega_{k\lambda}=6.84,7.846$ eV, that formed during the laser pulse action (see fig.\ref{Fig4c}). To provide more insight into the nature of these peaks we present the distribution of the values $\left| C_{n}^{(k\lambda)} \right|^2$ over $n$ for these two frequencies (see fig.\ref{Fig5}). It was found that total probabilities $\sum\limits_{n} {\left| C_{n}^{(k\lambda)} \right|^2}$ to find the atom in the bound state after the emission of photon were 0.355 and 0.465 for $\hbar\omega_{k\lambda}=6.84$  and $7.84$ eV respectively. It means that these peaks  appear dominantly as a result of a bremsstrahlung effect. Nevertheless, significant part of this quanta appears from the transition to Rydberg states with $n=5,7,9$. These transitions can result from hyper-Raman scattering with absorption of a number of photons of laser field and spontaneous emission of photon $\hbar\omega_{k\lambda}$. As these peaks are formed during the laser pulse action the AC Stark shift of different levels should be taken into account. As far as for the continuum boundary and Rydberg states the Stark shift can reach several eV, the number of absorbed laser photons is probably $5$ or $7$.
\begin{figure}
\centering\includegraphics[width=8 cm]{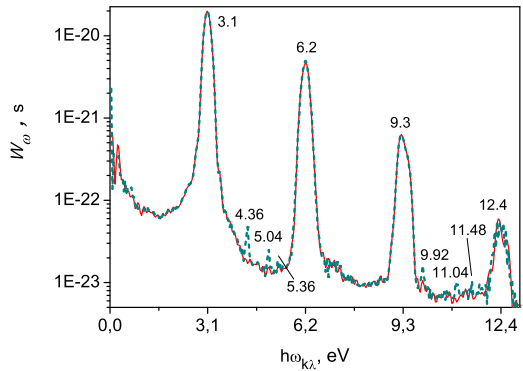}
\caption{{\protect\footnotesize {Similar to fig.\ref{Fig3}, but for laser intensity $10^{15}$ W/cm$^2$.}}}\label{Fig6}
\end{figure}
    
     To conclude our study we will discuss the emission for extremely high laser intensity $10^{15}$ W/cm$^2$. The spectrum of emission for this intensity is presented at Fig.\ref{Fig6} and consists of both odd and even harmonics of laser radiation that are formed during the laser pulse action. Several peaks of spontaneous transitions emerging after the laser pulse are well pronounced and can also be considered as the manifestation of stabilization phenomenon. Among them there are peaks corresponding to transitions $\ket{n=5,7,9}\to\ket{2}$ as well as new series $\ket{n=4,6,8}\to\ket{1}$ with lines $9.92, 11.04, 11.48$ eV.
     \begin{figure}
	\centering\includegraphics[width=8 cm]{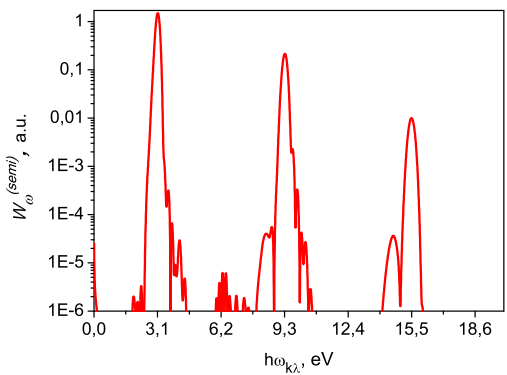}
	\caption{{\protect\footnotesize {Probability of emission of an atom irradiated by the 2nd harmonic of Ti-Sa laser pulse with peak intensity $3\times10^{13}$ W/cm$^2$ (semiclassical approach).}}}\label{Fig7}
	\end{figure}
     
     However the dominant part of atomic emission comes from laser frequency harmonic generation. Preliminary analysis of this emission leads to the conclusion that observed harmonics appear from the bremsstrahlung effect, but not from photorecombination. This fact is in contradiction with semiclassical approach \cite{fedorov1997atomic} typically applied to study the harmonic generation process in strong laser fields. This approach \cite{l1993high,lewenstein1994theory,becker1994modeling,platonenko1998generation,krause1992high} is based on the calculation of the averaged over the atomic wave functions dipole moment of the quantum system:
	\begin{equation}
	\vec{d}(t)=-e\int|\psi(\vec{r},t)|^2 \vec{r}d^{3}r,
	\label{dipole_mom} 
	\end{equation}	 		            
under the assumption that the spectral intensity of emission is proportional to the the Fourier transform of squared second derivative of (\ref{dipole_mom}). Than the semiclassical probability of emission can be found as $W_\omega^{(semi)} \sim \omega^3 \abs{\vec{d}_\omega}^2$. 

Generally, this semiclassical model is absolutely inconsistent with the approach based on the study of atomic system interaction with quantized field modes \cite{bogatskaya2016spontaneous}. The main contradiction appears from the fact that transitions only to the populated states can take place in the semiclassical model \cite{bogatskaya2016polarization}. Hence, almost all spectral lines that appear in our simulations can not be observed in the semiclassical approach.
   
To illustrate this fact we provide here the results of semiclassical calculations of spontaneous emission probability $W_\omega^{(semi)}$(in arbitrary units) of the model atom in the presence of the laser pulse with intensity $3\times 10^{13}$ W/cm$^2$ (fig.\ref{Fig7}). These data should be compared with those presented at fig.\ref{Fig4a}. First, we would like to stress the absence of the emission corresponding to the transitions to the low-lying bound states. This problem was discussed in \cite{bogatskaya2016polarization} and results from the fact that these states are unpopulated during the pulse and in the after-pulse regime. Second, we can see that only odd harmonics of fundamental frequency can be emitted in semiclassical approach while quantum electrodynamical considerations allows even harmonics of laser radiation as well. 

\section{Conclusion}

Thus, general approach to analyze the spontaneous emission of an atomic system driven by a strong laser field was discussed in detail. It is based on the first order of perturbation theory for the interaction with quantized vacuum field modes while the interaction with the intense classical laser field is considered numerically or analytically beyond the perturbation theory. The spontaneous emission from the model one-dimensional atom driven by the high-intensity field of second harmonic of Ti-Sa laser was studied. Different types of transitions between discrete and continuum states were found to exist and distinguished. It was demonstrated the spontaneous emission from the atom in the after-pulse regime can be used to testify the population trapping in atomic Rydberg states and interference stabilization phenomenon. 
    
   We have also demonstrated that data for emission spectra of a single atom obtained in the frames of semiclassical approach are found to be absolutely inconsistent with the results of quantum-electrodynamical calculations. This conclusion can be very important for understanding of physics of high order harmonic generation and should be studied in more detail. 
    
\section*{Acknowledgement}

This work was supported by the Russian Foundation for Basic Research (projects no. 15-02-00373, 16-32-00123). Numerical modeling was performed on the Lomonosov MSU supercomputer. Authors thank M.V.Fedorov for fruitful discussion.

\bibliographystyle{unsrtnat}
\bibliography{text1}

\end{document}